\documentclass[11pt]{article}
\usepackage{apalike,latexsym,amssymb,bbm,newlfont}

\def\titel{The Principles of Gauging}

\def\ben{\begin{enumerate}}
\def\een{\end{enumerate}}
\def\bi{\begin{itemize}}
\def\ei{\end{itemize}}
\def\bd{\begin{description}}
\def\ed{\end{description}}
\def\bq{\begin{quote}}
\def\eq{\end{quote}}
\def\bc{\begin{center}}
\def\ec{\end{center}}
\def\be{\begin{equation}}
\def\ee{\end{equation}}

\def\setP{\mathbbm P}
\def\setR{\mathbbm R}
\def\Mink{\setR^{(1,3)}}              
\def\DiffM{{\cal D\mbox{\em iff}(M)}} 
\def\AutP{ {\cal A\mbox{\em ut}(M)} } 

\def\semisub{{\times \hspace{-1.1em} \subset}}

\def\ket#1{ | #1 \rangle }

\def\braket#1#2{ \langle #1 | #2 \rangle }

\renewcommand{\thefootnote}{\arabic{footnote}}
\renewcommand{\baselinestretch}{1.1}
\begin{document}
\sloppy

\thispagestyle{empty}

\bc

\vspace*{5mm}

{

\renewcommand{\thefootnote}{\fnsymbol{footnote}}

{\huge {\Huge \titel}\footnote{Talk at PSA2K meeting
in Vancouver, B.C., November 4-7, 2000.
To be published in {\em Philosophy of Science}.
}}

\vspace*{8mm}

{\Large Holger Lyre}\footnote{Institut f\"ur Philosophie,
Ruhr-Universit\"at Bochum, D-44780 Bochum, Germany,
Email: holger.lyre@ruhr-uni-bochum.de
}\footnote{Special thanks to Tim Oliver Eynck for helpful remarks.}

\setcounter{footnote}{0}

}
{\footnotesize Ruhr-University Bochum}

\ec

\vspace*{8mm}

\hrule height 1pt

\bigskip

{

\renewcommand{\baselinestretch}{1} \large \footnotesize

{
\noindent The aim of this paper is twofold:
First, to present an examination of the principles underlying
gauge field theories. I shall argue that there are two principles
directly connected to the two well-known theorems of Emmy Noether
concerning global and local symmetries of the free matter-field
Lagrangian, in the following referred to as
``conservation principle'' and ``gauge principle''.
Since both these express nothing but certain symmetry features
of the free field theory, they are not sufficient to derive
a true interaction coupling to a new gauge field.
For this purpose it is necessary to advocate a third,
truly empirical principle which may be understood
as a generalization of the equivalence principle.
The second task of the paper is to deal with the ontological question
concerning the reality status of gauge potentials in the light
of the proposed logical structure of gauge theories.
A nonlocal interpretation of topological effects in gauge theories
and, thus, the non-reality of gauge potentials in accordance with
the generalized equivalence principle will be favoured.
}

}

\bigskip

\hrule height 1pt

\vspace{8mm}


\paragraph{1. The Gauge Argument.}
Textbook presentations of the logical structure of gauge field
theories usually emphasize the importance of the
{\em gauge principle} (cf. Aitchison and Hey 1982, p. 176):
Start with a certain free field theory---take Dirac's theory
${\cal L}_D = \bar\psi (i \gamma^\mu \partial_\mu - m ) \psi$
for instance---and consider local gauge transformations
\be
\psi(x) \to \psi'(x) = e^{iq \alpha(x)} \psi(x) .
\ee
To satisfy the requirement of local gauge covariance of ${\cal L}_D$
the usual derivative has to be replaced by a covariant derivative
\be
\label{cov_der}
\partial_\mu \ \to \ D_\mu = \partial_\mu - i q A_\mu .
\ee
Thus, instead of a free field theory,
we obtain---in a somewhat miraculous way---a theory with interaction
\be
{\cal L}_D \ \to \
{\cal L}_D' = \bar\psi (i \gamma^\mu D_\mu - m ) \psi
= \bar\psi (i \gamma^\mu\partial_\mu - m) \psi + q \bar\psi \gamma^\mu
A_\mu \psi .
\ee

Besides this, in usual textbook presentations reference is also made
to the importance of ``Noether's theorem'':
The existence of a global (i.e. rigid) $k$-dimensional symmetry group
is connected with the existence of $k$ conserved currents.
This general result may very well be referred to as a principle of
its own, here called the {\em conservation principle}.
In the case of Dirac's theory we find that ${\cal L}_D$ exhibits
global gauge covariance under
$\psi(x) \to \psi'(x) = e^{iq \alpha} \psi(x)$.
The Noether current then reads
$\jmath^\mu(x) = - q \bar\psi(x) \gamma^\mu \psi(x)$.
Now, the ``miracle'' of the gauge principle consists in the idea
that by simply postulating local gauge covariance one is led to
introduce a new interaction potential $A_\mu(x)$ obeying
$A_\mu(x) \to A'_\mu(x) = A_\mu(x) - \partial_\mu \alpha(x)$.
Surely, it is very tempting to hold this suggestive interpretation,
but are we really forced to?

From a mathematically more rigorous point of view,
the above presentation is a bit to much of a ``miracle''.
Both the conservation principle as well as the gauge principle
are simply concerned with Noether's first and second theorem--and
thus with a mere symmetry analysis of the free field theory.
Let $\phi_i(x)$ be a field variable and let $i$ be the index of
the field components, then Noether's first theorem states
that the invariance of the action functional
$S[\phi]=\int {\cal L} [ \phi_i(x), \partial_\mu \phi_i(x) ] \, d^4x$
under the action of a $k$-dimensional Lie group implies
the existence of $k$ conserved currents.
This is what is usually just called ``Noether's theorem''.
In the language of the underlying fiber bundle structure,
Noether's first theorem points to the importance
of the bundle structure group---in the case of the above Dirac-Maxwell
theory the gauge group $G=U(1)$. Hence, we are working with
a $U(1)$-principal bundle $\setP$ over spacetime.\footnote{A more
detailed presentation of gauge theories and their bundle structure
as well as a brief account of the theory of fiber bundles
can be found in Guttmann and Lyre (2000).}

Now let ${\cal G} = \AutP \simeq \DiffM \semisub G$
be the automorphism group of $\setP$.
Noether's second theorem, then, states that the invariance of
the action $S[\phi]$ under $\cal G$ implies
the existence of $k$ constraints known as Bianchi identities.
From (\ref{cov_der}) we first of all find the Jacobi identity
$\epsilon^{\mu\nu\rho\sigma} [D_\nu, [D_\rho, D_\sigma] ] = 0$.
With the definition $F_{\mu\nu} = \partial_\mu A_\nu - \partial_\nu A_\mu$
this is equivalent to the Bianchi identity
$\epsilon^{\mu\nu\rho\sigma} D_\nu F_{\rho\sigma} = 0$.
Clearly, ${\cal G}$ is infinite-dimensional and consists of
spacetime-dependent (i.e. ``local'') group elements.
In this way Noether's second theorem gives rise to the postulate
of local gauge covariance which in turn underlies the gauge principle.

Note again that Noether's analysis is just concerned with
the symmetry conditions of a given action functional.
Hence, this does not allow for---or even force us---
to interpret the $A_\mu$-term as a new interaction term.
Both the conservation as well as the celebrated gauge principle
lay claim to certain symmetry conditions of a given field
theory---without introducing a new field.
Indeed, how could a new physical field be derived from a mere
analysis of the symmetry structure of some theory?
How, then, are we to understand the occurence of the $A_\mu$-term?


\paragraph{2. Intrinsic Gauge Theoretic Conventionalism.}
In recent times, a critical reading of the gauge argument
has been presented by several authors;
cf. Brown (1999), Healey (2000), Teller (2000).
To easily see the issue consider a wavefunction
$\Psi(x) = \braket{x}{\phi}$ in the position representation $\ket{x}$
(with $\Big\{\ket{\phi}\Big\}$ spanning an abstract Hilbert space).
Now, local gauge transformations read
$\ket{x} \ \to \ \ket{x'} = e^{i \alpha(x)} \ket{x} = \hat U \ket{x}$.
Such a transformation acts as changing the representation
basis of the Hilbert space and, thus,
operators on that Hilbert space have to be transformed, too.
A general operator transformation looks like
$\hat O' = \hat U \hat O \hat U^+$.
In the particular case of the derivative operator we find
$\partial_\mu \to D_\mu = \partial_\mu - iq A_\mu(x)$
with the definition $A_\mu(x) = - \partial_\mu \alpha(x)$.

We therefore clearly see that (\ref{cov_der}) has to be understood
as a mere change in the position representation expressed in terms
of local gauge transformations.
Hence, the covariant representation of the derivative
is as conventional as a mere coordinate representation.
This feature might very well be called an
``intrinsic gauge theoretic conventionalism''.
The clear consequence of this is that the celebrated gauge principle
is not sufficient to derive the coupling to a new interaction-field.
No new physics enters, no new physical field is really introduced!


\paragraph{3. A Missing Principle.}
A true gauge field theory should be considered as a coupling
between a matter-field and an interaction-field theory.\footnote{For
an elaboration of the next two sections the reader may want to refer
to my companion paper Lyre (2001).}
We are therefore faced with the following problem.
We have, on the one hand, equations of motion
of the free matter-field (e.g. Dirac's equation).
Due to the gauge principle the Lagrangian reads
\be
\label{dirac-inhom}
{\cal L}_D' = {\cal L}_D + {\cal L}^{(i)}_{inhom}
= \bar\psi (i \gamma^\mu\partial_\mu - m) \psi - \jmath^{(i)}_\mu A^\mu
\ee
with an unphysical inhomogeneity term, since the connection field
$A^\mu$ is flat (i.e. the curvature gauge field vanishes).
On the other hand we have certain gauge field equations
(Maxwell or Yang-Mills equations)
\be
{\cal L}_F' = {\cal L}_F + {\cal L}^{(f)}_{inhom}
= - \frac{1}{4} F_{\mu\nu} F^{\mu\nu} - \jmath^{(f)}_\mu A^\mu .
\ee
Here, the inhomogeneity stems from the field sources,
i.e. certain ``field charges'' $q^{(f)}$.
In contrast to this the vector current $\jmath^{(i)}$
in (\ref{dirac-inhom}) implies a factor $q^{(i)}$ which is
due to the phase $e^{iq^{(i)} \alpha}$ of the Dirac wavefunction.
As will be explained in a moment we will call this
the ``inertial charge''.

In (\ref{dirac-inhom}), the inhomogeneity term
${\cal L}^{(i)}_{inhom}$ stems from the gauge principle.
It should be clear from our critical reading of this principle
that there is no {\em a priori} possibility to identify
${\cal L}^{(i)}_{inhom}$ and ${\cal L}^{(f)}_{inhom}$,
or $q^{(i)}$ and $q^{(f)}$, respectively.
Since both conservation principle as well as gauge principle
turned out as mere analytic statements about the symmetry structure
of the free matter-field theory, we are in need of a truly
empirical---synthetic so to speak---principle of gauging,
which allows for the identification of
${\cal L}^{(i)}_{inhom}$ and ${\cal L}^{(f)}_{inhom}$.
Fortunately, the gauge theoretic analogy to general relativity
may help to find such a missing principle.

In fact, in standard general relativity we may also formulate
a gravitational gauge principle.
The starting point for this is the free geodesic equation
$\frac{d}{d \tau} \theta^\mu_\alpha(\tau) = 0$
for a tetrad reference frame $\theta^\mu_\alpha$.
The gauge principle demands covariance under local $SO(1,3)$ or
$\Mink$ transformations.\footnote{It depends on the Noether current
arising from the conservation principle how to couple the gravitational
field. Certainly, a straightforward choice for the gauge group
of general relativity is the group of Poincar\'e translations $\Mink$
which implies the conservation of energy-matter.
Moreover, this is a reasonable choice since local translations
are equivalent to diffeomorphisms.}
We get a covariant derivative
\be
\frac{d}{d \tau} \theta^\mu_\alpha(\tau) \
\to \ \nabla_\tau \theta^\mu_\alpha(\tau)
= \frac{d}{d \tau} \theta^\mu_\alpha(\tau) + \Gamma^\beta_{\nu\alpha}
\frac{d x^\nu(\tau)}{d \tau} \theta_\beta^\mu(\tau) .
\ee
Now, the Levi-Civita connection $\Gamma_\mu$ does
not necessarily represent a true gravitational potential
with a non-vanishing gravitational field (i.e. Riemann curvature).
Indeed, $\Gamma_\mu$ might occur simply because of a peculiar
choice of coordinates!

The ``true'' gravitational field with non-vanishing Riemann
curvature is of course governed by the Einstein field equations
$R_{\mu\nu} - \frac{1}{2} \, R \ g_{\mu\nu} = - \kappa \ T_{\mu\nu}$.
The r.h.s. represents the field source, i.e.
gravitational mass $m^{(g)}$ which is encoded in the
energy-momentum tensor $T_{\mu\nu}$.
In contrast to this, a freely moving observer in spacetime
(a reference frame represented by a tetrad in the geodesic
equation) will be assigned an inertial mass $m^{(i)}$.
As the reader might guess already, in general relativity
the conceptual problem of linking equations of motion
and field equations is solved on the basis of the
{\em equivalence principle}. The cruicial identification
\be
\label{mimf}
m^{(i)} = m^{(g)}
\ee
becomes indeed the one and decisive empirical input
to any geometric theory of gravitation.
There is no {\em a priori} reason to identify inertial
and gravitational mass and, hence, the equivalence principle
has to be vindicated by the experimental fact that different
materials do have the same free fall behaviour.
In this way, the universality of the gravitational coupling
constitutes a deep fundamental insight.


\paragraph{4. The Generalized Equivalence Principle.}
We may now return to our problem of finding the missing empirical
principle in gauge field theories.
The relativistic equivalence principle implies the possibility
of a non-flat connection and, hence, a non-vanishing gravitational field.
Due to the close analogy between general relativity and standard
quantum gauge field theories we may very well {\em generalize}
the idea of the equivalence principle.
Indeed, the equivalence of inertial and field charges
\be
\label{qiqf}
q^{(i)} = q^{(f)}
\ee
solves our problem and allows for a true coupling term
${\cal L}_{coup}={\cal L}^{(i)}_{inhom}={\cal L}^{(f)}_{inhom}$.
Thus, equations of motion and field equations belong to one combined
framework and we obtain the full Lagrangian of a gauge field theory
representing---quite generally---the coupling between matter-field
and interaction-field
\be
{\cal L}_{GFT}={\cal L}_D + {\cal L}_{coup} + {\cal L}_F .
\ee

We may give the following geometric formulation
of the {\em generalized equivalence principle}:
\bq
{\bf GEP:}
{\em It is always possible to perform a local gauge transfomation
such that, locally (i.e. at a point), the gauge field vanishes.}
\eq
In this way, GEP implies a non-flat connection, i.e. a gauge potential
which is irrevocably connected with the occurrence of an
interacting gauge field originating in the field charges
and obeying its own dynamics.
Equation (\ref{qiqf}) turns out as a direct consequence of this,
for if we regard the connection as non-flat,
the field must have its sources in certain field charges.
Moreover, GEP includes the interaction-free theory
as a local limiting case.

The reader may wonder whether we have simply replaced one miracle
by another. However, the equivalence (\ref{qiqf}) is far from trivial.
There is---quite analogous to (\ref{mimf})---no {\em a priori} reason
to identify inertial and field charges.
Let us assume for a moment $\frac{q^{(f)}}{q^{(i)}} \ne 1$.
This means that different types of particles of equal electric charge
would couple differently to the electromagnetic field.
We should expect a difference in the coupling of electrons and muons
or d-quarks and s-quarks---to give but two examples---and should
therefore write down different Dirac equations
\be
(i \gamma^\mu \partial_\mu - m ) \psi e^{i q^{(i)} \alpha}
= c_p \ q^{(f)} \ \gamma^\mu A_\mu \ \psi e^{i q^{(i)} \alpha}
\ee
for different types of particles with the same $q^{(f)}$
but a particle-type dependent factor $c_p$.
This is clearly not what we observe.
In fact, GEP predicts a whole variety of {\em null-experiments}
(as does its relativistic counterpart).
The equivalence (\ref{qiqf}) indicates the empirically known
universality of the gauge field coupling, turning GEP into
the one and decisive physical principle of gauge theories.


\paragraph{The three principles of gauging}

We may summarize our considerations so far.
It will be helpful to draw the following distinction
of types of fiber bundles occuring in gauge theories:
We may have trivial bundles\footnote{Trivial bundles
allow for global sections, they globally look like direct products
spaces. In contrast to this, non-trivial bundles only locally look
like a direct product.}
with flat and non-flat connections.
Let us call them type 1 and type 2 bundles.
As long as we are concerned with trivial bundles,
the notion of a fiber bundle is in a way superfluous
(since we may simply use a direct product).
For non-trivial bundles, however, the fiber bundle framework
becomes indispensible.
Let us indicate non-trivial bundles as type 3 bundles.
Since we may again distinguish between flat and non-flat connections,
we may accordingly call them type 3a and type 3b bundles.

I shall rewiew the three proposed principles of gauging:
\bd
\item[Conservation principle.]
Based on Noether's first theorem it connects
the global (i.e. rigid) symmetry of a free field theory
with the existence of certain conserved quantities.
As an analytic statement of the symmetry structure of the theory
it does not contain any new physical information.

\item[Gauge principle.]
Based on Noether's second theorem it connects
the local (i.e. spacetime-dependent) symmetry of a free field
theory with the {\em suggested} structure of the coupling to an
interaction-field. It is tempting to take the suggested coupling already
for granted, however, the gauge principle only implies flat connections
and, hence, no non-vanishing interaction fields.
In other words, the gauge principle does not allow for a transition
from type 1 to type 2 bundles (or type 3a to 3b, respectively).
As a mere analytic statement of the symmetry structure of the theory
it also does not contain new empirical information.

\item[Equivalence principle.]
This is a true empirical principle which lays claim
for the universality of the gauge field coupling
due to the identification $q^{(i)} = q^{(f)}$.
This manifests the coupling of matter-fields and
interaction-fields and allows to combine equations of motion
and field equations into one framework.
The equivalence principle implies the existence of non-flat connections
and therefore non-vanishing interaction-fields.
It is a synthetic statement of the empirical basis of gauge field
theories.

\ed

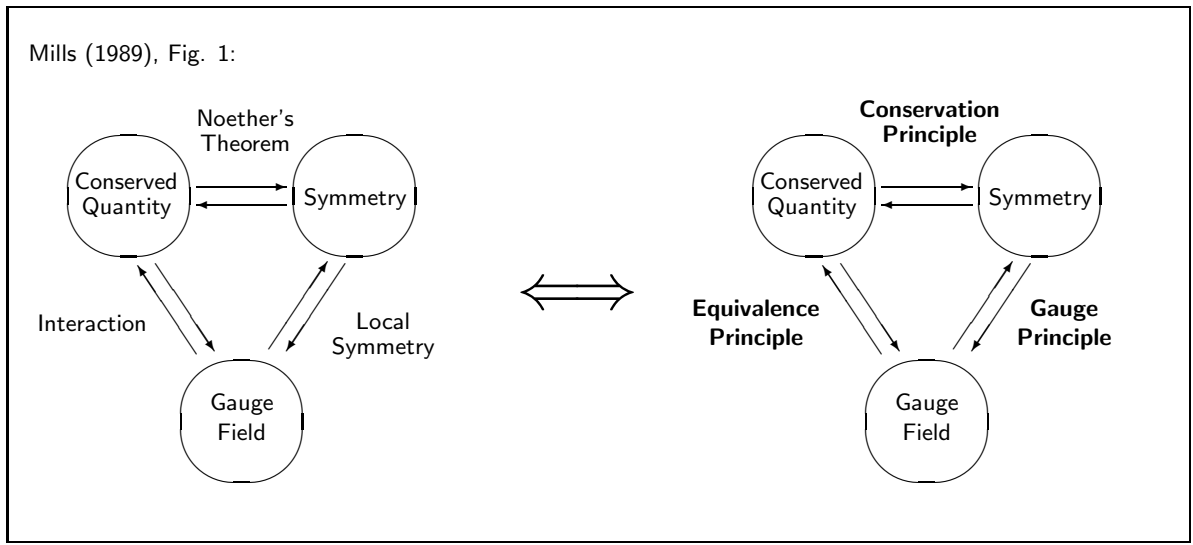
\begin{figure}[t]
\bc
\fbox{\parbox{155mm}{
\footnotesize\sf
\unitlength.6mm
\begin{picture}(150,115)                 

\put(1,105){ \footnotesize Mills (1989), Fig. 1: }

\put(25,75){\oval(27,27)}\put(13,71){\shortstack[c]{Conserved\\Quantity}}

\put(40,85){\shortstack[c]{Noether's\\Theorem}}
\put(40,77){\vector(1,0){20}}
\put(60,73){\vector(-1,0){20}}

\put(75,75){\oval(27,27)}\put(62,73){ Symmetry }

\put(70,40){\shortstack[c]{Local\\Symmetry}}
\put(73,60){\vector(-2,-3){13}}
\put(56,41){\vector(2,3){13}}

\put(50,25){\oval(27,27)}\put(43,21){\shortstack[c]{Gauge\\Field} }

\put(3,45){ Interaction }
\put(40,40){\vector(-2,3){13}}
\put(31,60){\vector(2,-3){13}}

\put(111,50){\Huge $\Longleftrightarrow$}
\end{picture}
\begin{picture}(100,100)        

\put(25,75){\oval(27,27)}\put(13,71){\shortstack[c]{Conserved\\Quantity}}

\put(35,87){\shortstack[c]{\bf Conservation\\ \bf Principle}}
\put(40,77){\vector(1,0){20}}
\put(60,73){\vector(-1,0){20}}

\put(75,75){\oval(27,27)}\put(62,73){ Symmetry }

\put(70,42){\shortstack[c]{\bf Gauge\\ \bf Principle}}
\put(73,60){\vector(-2,-3){13}}
\put(56,41){\vector(2,3){13}}

\put(50,25){\oval(27,27)}\put(43,21){\shortstack[c]{Gauge\\Field} }

\put(-2,42){\shortstack[c]{\bf Equivalence\\ \bf Principle}}
\put(40,40){\vector(-2,3){13}}
\put(31,60){\vector(2,-3){13}}

\end{picture}
}}
\ec
{\sf
\centerline{
\parbox{110mm}{
\caption{\label{mills-pattern}
In his 1989 review article, Robert Mills gave a graphical representation
of what he calls the ``logical pattern of a gauge theory''.
The triangular structure of his pattern, on the left hand side,
also illustrates our understanding of the principles of gauging
in terms of the right hand side figure.
}}}}
\end{figure}

A schematic representation is given in figure \ref{mills-pattern}.

\paragraph{6. The Reality of Gauge Potentials.}
``Only gauge-independent quantities are observable.''
This truism is supported by our critical remarks on the
intrinsic gauge theoretic conventionalism of local symmetries.
It is also in accordance with GEP as an argument in favour of non-flat
connections, i.e. non-vanishing gauge-independent field strengths.
Therefore, GEP lays claim for not considering gauge potentials
as physically real entities.
Clearly, this is true for type 2 and type 3b bundles
which are concerned with non-flat connections.
However bundles of type 3a seem to allow for
physical---viz. topological---effects
which have their origin in flat connections
(type 1 is just a trivial case).
Does this contradict GEP's point of view of not considering
gauge potentials as physically real?

Usually physicists think along these lines.
They do consider gauge potentials as real entities
because of topological effects in field theories.
This view is supported by some kind of a common-sense
indispensability argument:
First, gauge potentials---and matter-fields---are the genuine objects
in the fiber bundle formulation of gauge theories.
They are clearly indespensible for the mathematical formulation
(as being the connection forms).
Also, they are indespensible for the physical formulation
of quantum field theories, since
both the {\em coupling structure} (vertex structure) as well as
the {\em quantization procedure} itself are represented
on the level of potentials and not the field strengths.
How, then, are we to do physics without potentials?

As Michael Redhead (2000) has pointed out, the situation is even worse,
since no matter wether we consider potentials real or not,
we will always face ontological problems.
Quite generally, such problems seem to arise in theories
with a certain mathematical {\em surplus structure}.
Here we have a mathematical structure $M'$ which is larger
than the structure $M$ needed for a direct correspondence
(i.e. isomorphism) to the observable physical structure $P$.
The complement of $M$ in $M'$ might be called surplus structure.
Gauge potentials are an example of surplus structure in gauge theories.
Now, ``Redhead's dilemma'' looks like the following:
On the one hand the reality of gauge-dependent potentials
implies a mystic influence from non-observable physical beables
to observable ones. This, in fact, is a version of the famous
hole argument---and this first horn of the dilemma
leaves us with indeterminism.\footnote{For a discussion of
the bundle space hole argument see Lyre (1999).}
The second horn is that once we assert the non-reality
of gauge potentials, this implies a ``Platonist'' role
for mathematical elements to influence physical beables.

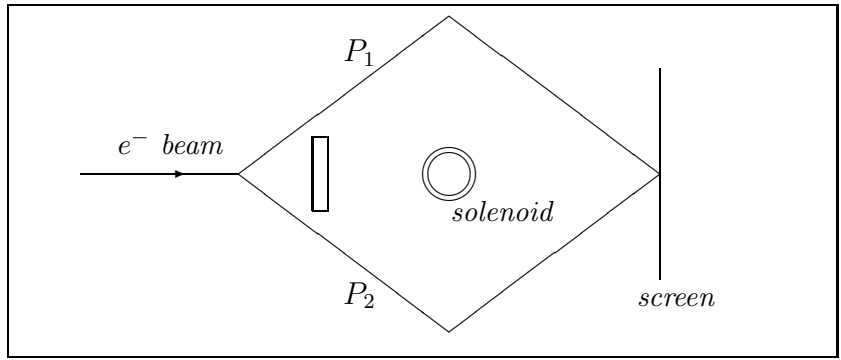
\begin{figure}[t]
\bc
\fbox{\parbox{108mm}{
\unitlength.7mm
$$
\begin{picture}(110,60)

\put(7,36){\makebox(0,0){\em $e^-$ beam}}
\put(0,30){\vector(1,0) {10}}
\put(10,30){\line(1,0) {10}}
\put(-10,30){\line(1,0) {10}}

\put(34,23){\line(0,0) {14}}
\put(34,23){\line(1,0) {3}}
\put(34,37){\line(1,0) {3}}
\put(37,23){\line(0,0) {14}}

\put(10,30){\line(1,0) {10}}

\put(20,30){\line(4,3) {40}}
\put(20,30){\line(4,-3){40}}
\put(60,60){\line(4,-3){40}}
\put(60,0){\line(4,3) {40}}
\put(43,53){\makebox(0,0){\large $P_1$}}
\put(43,7){\makebox(0,0){\large $P_2$}}

\put(100,10){\line(0,0) {40}}
\put(103,6){\makebox(0,0){\em screen}}

\put(60,30){\circle{8}}
\put(60,30){\circle{10}}
\put(70,23){\makebox(0,0){\em solenoid}}

\end{picture}
$$
}}
\ec
{\sf
\centerline{
\parbox{110mm}{
\caption{\label{AB}
Schematic experimental configuration of the AB effect.
}}}}
\end{figure}

To get along with Redhead's dilemma we shall take a closer
look to the well-known Aharonov-Bohm effect,
which is indeed the paradigm case of type 3a bundles---and,
hence, topological effects.
Due to Aharonov and Bohm (1959) a shift
in the interference pattern of the electron wave function
surrounding a solenoid (on paths $P_1$ and $P_2$) is observed,
even though the electron is shielded from the region
of the magnetic field (see figure \ref{AB}).
Since only the vector potential has a non-vanishing
contribution outside the solenoid,
the AB effect is usually understood as showing the
physical significance of gauge potentials.

As can be seen from the experimental configuration,
the existence of the AB effect depends crucially
on the fact that the configuration space of the electron
is not simply-connected. Since the electron is shielded
from the solenoid, this space has essentially the toplogy
of a circle (as represented by any closed loop surrounding the solenoid,
such as paths $P_1$ and $P_2$, for instance).
Now, as Yang (1974) has first pointed out,
the AB effect may be described solely in terms
of the Dirac phase factor
\be
\Delta \alpha = \oint_{\cal C} A_\mu \ dx^\mu .
\ee
This integral lives in the space of loops and is called a holonomy.
Clearly, holonomies are gauge-independent quantities and
therefore appropriate candidates of observable entities.

So far, this does not solve our problem since we are still
working with an integral which depends on the gauge potential
(in a gauge-independent manner, though).
However, due to Stokes' theorem
\be
\oint_{\cal C} A_\mu \ dx^\mu
= \int_{\cal S} F_{\mu\nu} \ ds^{\mu\nu},
\ee
we might very well describe the AB effect as a {\em nonlocal effect}
in terms of the magnetic field strength alone.
In fact, Stokes' formula allows to shift back and forth
between the potential and the field strength interpretation.
Presented this way, the AB effect turns out as a nice case study
of theory underdetermination by empirical evidence.
Physicists tend to favour the potential interpretation
since it apparently allows for a local interaction account.
This, however, leaves Redhead's dilemma unsolved.

A second, even stronger worry against the physicist's common line
of simply accepting the reality of gauge potentials,
is the fact that, in any case,
due to the topological origin of Dirac's phase factor,
we will never completely get rid of a certain kind of
non-locality---or non-separability (Healey 1997).
Topological effects unavoidably lead, in one way or the other,
to a nonlocal account.
It is therefore impossible to give a purely local description
of the interference shift,
neither within the field nor the potential interpretation.
We may gladly accept the field interpretation and, also,
should consider holonomies as physically real.
This option has the clear advantage of avoiding Redhead's dilemma,
since no surplus structure arises.


\paragraph{7. Conclusion.}
Even for the description of toplogical effects in type 3a bundles,
the reality of gauge potentials is not enforced.
We may very well represent the physically significant structures
in an ontological universe consisting of matter-fields,
gauge field strengths and holonomies.
The price we pay is to accept a certain type
of nonlocality in gauge theories---which seemingly differs
from quantum nonlocalities such as EPR correlations
due to its manifest topological origin,
but seems unavoidable anyhow in both the potential
and the field strength account.

Now, since holonomies may be represented in terms of
gauge field strengths (because of Stoke's formula),
we are in perfect agreement with GEP as an argument
against the significance of flat connections.
Indeed, the three proposed principles of gauging prove to be
a consistent framework of the main conceptual structure
of gauge field theories.
Maybe, therefore, the idea of a generalized equivalence principle
helps to clarify the issue of the logical gauge theoretic pattern.
Nevertheless, a couple of deep philosophical puzzles remain
to be solved---last but not least the very idea of gauging itself,
which may heavily lean on a sufficient account of locality
and nonlocality in physics.
Thus, the issue of gauge theories should become much more
the focus of philosophers of science than it was before.
Personally, I couldn't agree more to how
Michael Redhead (2001) has recently put it:
{\em ``The gauge principle is generally regarded as
the most fundamental cornerstone of modern theoretical physics.
In my view its elucidation is the most pressing problem in current
philosophy of physics.''}


\newpage

\bibliographystyle{apalike}


\end{document}